\documentstyle[11pt,paspconf,epsf]{article}
\begin{document}
{\scriptsize \noindent Invited talk to be published in {\sc The
Evolution of Galaxies on Cosmological Timescales} \\
1999, eds. J.E. Beckman, \& T.J. Mahoney, ASP Conf. Ser.  
\vskip -1truemm}
\title{Birth, Aging, and Death of Galactic Bars} 
\author{Daniel Friedli} 
\affil{Observatoire de Gen\`eve, CH-1290 Sauverny, Switzerland;
  Daniel.Friedli@obs.unige.ch}

\begin{abstract}
  The life of disk galaxies is likely punctuated with the birth
  (spontaneous or induced), the aging (quick or slow), and the death
  (sudden or progressive) of bars. These events are responsible for
  various major changes of the galaxy which occur both on dynamical
  ($\sim$0.1\,Gyr) and cosmological ($\sim$10\,Gyr) time-scales, as
  well as both on large- ($\ga$10\,kpc) and small-scales
  ($\la$1\,kpc).  These modifications affect the morphology, the
  orbital structure, the dynamics, the star formation and central
  fueling rates, the abundance profiles, etc.  As a consequence,
  remarkable galaxy metamorphoses within the Hubble sequence can be
  observed.  The formation and the fate of gaseous bars are also
  briefly examined.
\end{abstract}


\section{Introduction}
The universe is not rigid or frozen. It is evolving, just like its
constituents; galaxies, stars, clusters are all changing with time.
Galactic {\it stellar bars} or any triaxial systems are not an
exception to the rule.  But they are not passively undergoing changes,
they are actively driving them (Sect.~2.2). Over the last thirty
years, mainly due to the development of efficient numerical
techniques, the required conditions for a bar to form have become
better known (Sect.~2.1).  Maybe the recent greatest surprise (but
certainly not the last one) about bars is that they can die as well
(Sect.~2.3).  As a consequence, unbarred galaxies can either not have
yet time to develop a bar, or have hosted one in the past which is now
dissolved.  An interesting question automatically arises: could a bar
be revived a few Gyr after it has been dissolved? Could bars even be
recurrent?  This is in principle possible, but it is certainly not a
straightforward task (e.g. Sellwood \& Moore 1999).  Anyway, $\beta$,
the ratio of disk galaxies having a stellar bar to the total number of
disk galaxies, is absolutely not supposed to be constant with time,
and should strongly vary with redshift $z$.  In what way? So far, the
answer is widely unknown and will remain so until high resolution deep
fields will routinely be available.

Gas distribution inside stellar bars is diverse and complex (Kenney
1997).  For instance, numerous {\it gaseous bars} are observed;
they however seem to be much less common than stellar bars.  This
might be due to an observational bias, but gaseous bars are certainly
not easy to form and/or have short lifetimes (Sect.~3).  Gas
morphology is clearly strongly dependent on the resolution and the
considered scale: a true bar-like shape (Maffei\,2) should not be
confused with either ring-like plus open arms (IC\,342), or ``twin
peaks'' (NGC\,3351) morphologies.

\section{Stellar Bars}
\subsection{Birth}
Stellar bars seem to be ubiquitous. At $z\approx0$ this is
indisputable since bars are found in $\sim$2/3 of bright disk galaxies
(Sellwood \& Wilkinson 1993; Knapen, this volume), and in $\sim$100\%
of Magellanic type galaxies (Odewahn 1996).  At higher $z$, there
seems to be very few bars as claimed by van den Bergh et al. (1996)
from visual classification of galaxies in HDF-north. But
observationally, it is certainly not yet possible to give accurate
numbers to $\beta(z)$.  First, because of the lack of resolution; for
instance at $z\!=\!1$, 0.1\arcsec\ corresponds to 1\,kpc, meaning that
only the largest bars could be detected even with HST or ground-based
facilities with adaptive optics. Second, because bars are best
detected in near-infrared (NIR) wavelengths, but very rarely in UV; at
$z\!=\!1$, observations in the visible (respectively NIR) correspond
in fact to UV (visible) in the galaxy rest frame and will highly
underestimate $\beta$.

{\it Spontaneous bars} are formed from unstable and nearly isolated
stellar disks (Hohl 1971; Sellwood 1981; Pfenniger \& Friedli 1991;
Fux 1997).  A convenient way to know if the disk will develop a bar is
to look at the Toomre (1964) parameter $Q_*$ for the stars.  The
condition for the instability onset is $Q_*\la2.0-2.5$ at all radii
(Athanassoula \& Sellwood 1986).  This is however true only when the
density gradient towards the galactic center is not too steep;
galaxies with dense centers can be stable with much lower $Q_*$
values.  In this case, the disk might instead suffer from $m\!=\!1$
instabilities.  Furthermore, supermassive black holes (SBH) can
completely prevent the $m\!=\!2$ mode from developing when reaching a
few percent of the stellar disk mass, even if $Q_* \approx 1.5$
throughout the disk (Friedli 1994).

{\it Induced bars} are triggered by close interactions (Noguchi 1988;
Gerin et al. 1990; Barnes \& Hernquist 1991). In this case, the
instability condition is slightly less severe $Q_*\la2.5-3.0$. Induced
bars can form in both lighter and hotter stellar disks. Typically,
spontaneous stellar bars should only appear $\sim$6\,Gyr after the
beginning of the disk build-up (Noguchi 1996), i.e. when
$z\approx0.5$.

According to Noguchi (1996) and Miwa \& Noguchi (1998), the properties
of spontaneous or induced bars strongly differ: spontaneous ones would
essentially form in late-type objects, show an exponential density
profile, and be fast-rotating. On the contrary, induced bars would be
manufactured in early-type galaxies, present a flat density profile,
and rotate slowly. However, this scenario remains to be confirmed;
indeed, it is for instance in contradiction with the claim by Combes
\& Elmegreen (1993) that late-type bars should be slow, i.e. end near
their Inner Lindblad Resonance (ILR).  Also, Kent (1987) and
Merrifield \& Kuijken (1995) showed that the bar of the early-type
galaxy NGC\,936 (SB0) is rotating very quickly.

\subsection{Aging}
Once they are born, stellar bars are not immutable. They are the scene
of numerous evolutionary processes which may affect the whole galaxy
(e.g. Martinet 1995).  These events occur on dynamical and
cosmological time-scales, as well as on large- and small-scales.
Below, four selected examples are briefly discussed.

\subsubsection{Evolution of Bar Pattern Speed.}
Since the orbital structure significantly depends on the bar pattern
speed $\Omega_p$, it is very important to know its precise value and
how it evolves.  For instance, a low $\Omega_p$ favors the onset of
ILRs and the anti-bar $x_2$, $x_3$ families of periodic orbits.
Various numerical simulations have clearly revealed that $\Omega_p$ is
generally not constant.

In purely collisionless $N$-body simulations, $\Omega_p$ decreases
with a typical time-scale $\tau_{\Omega}\approx5$\,Gyr (Combes \&
Sanders 1981; Little \& Carlberg 1991; Pfenniger \& Friedli 1991). The
decrease is fast during the first few rotations and then stabilizes at
a lower level.  This behavior can be explained by the presence of many
escaping chaotic particles, mainly originated from the corotation
radius $R_{\rm CR}$, which carry away significant angular momentum. A
decrease of $\Omega_p$ means that $R_{\rm CR}$ move outwards. But the
system is continuously re-adjusting so that the bar length $a$ also
increases.  Generally, the ratio ${\cal R} \!=\! R_{\rm CR}/a$ tends
to grow.  According to Elmegreen (1996; see also references therein),
early-type bars have ${\cal R} \approx 1.2$, and late-type ones ${\cal
  R} \approx 2$. However, Aguerri et al. (1998) found a weaker
dependence on morphological type (from ${\cal R} \approx 1.1$ to
${\cal R} \approx 1.4$); they showed as well that ${\cal R}$ slightly
increases with bar strength.

When a massive live dark matter halo (DMH) is present as well, the
slow down can be even more pronounced due to the dynamical friction
generated by the DMH on the bar (Fux et al.  1995; Debattista \&
Sellwood 1998).  $\tau_{\Omega}$ is inversely proportional to the mass
of the DMH and could be as small as a few hundreds of millions years.
Furthermore, direct and indirect determinations of $\Omega_p$ in
external galaxies show that many bars are likely fast-rotating. This
has thus led Debattista \& Sellwood (1998) to postulate that there
should only be a weak DM contribution in the bar region.

When a dissipative component is present or if the galaxy suffers from
a significant interaction, other behaviors may occur.  For instance,
the central bar-driven gas fueling suppresses the decrease of
$\Omega_p$ which then remains nearly constant or can even be
accelerated (Friedli \& Benz 1993; Berentzen et al. 1998). Gas loses
angular momentum in favor of the bar.  But this only is a relatively
modest and above all temporary effect, at long run $\Omega_p$
decreases or the bar is dissolved (see Sect.~2.3). Close and nearly
co-planar interactions may lead $\Omega_p$ to fluctuate by $\sim$10\%
(Gerin et al. 1990; Miwa \& Noguchi 1998). In brief, $\Omega_p$
slightly increases when the perturber leads the bar, and vice-versa.

\begin{figure}[t]
\plottwo{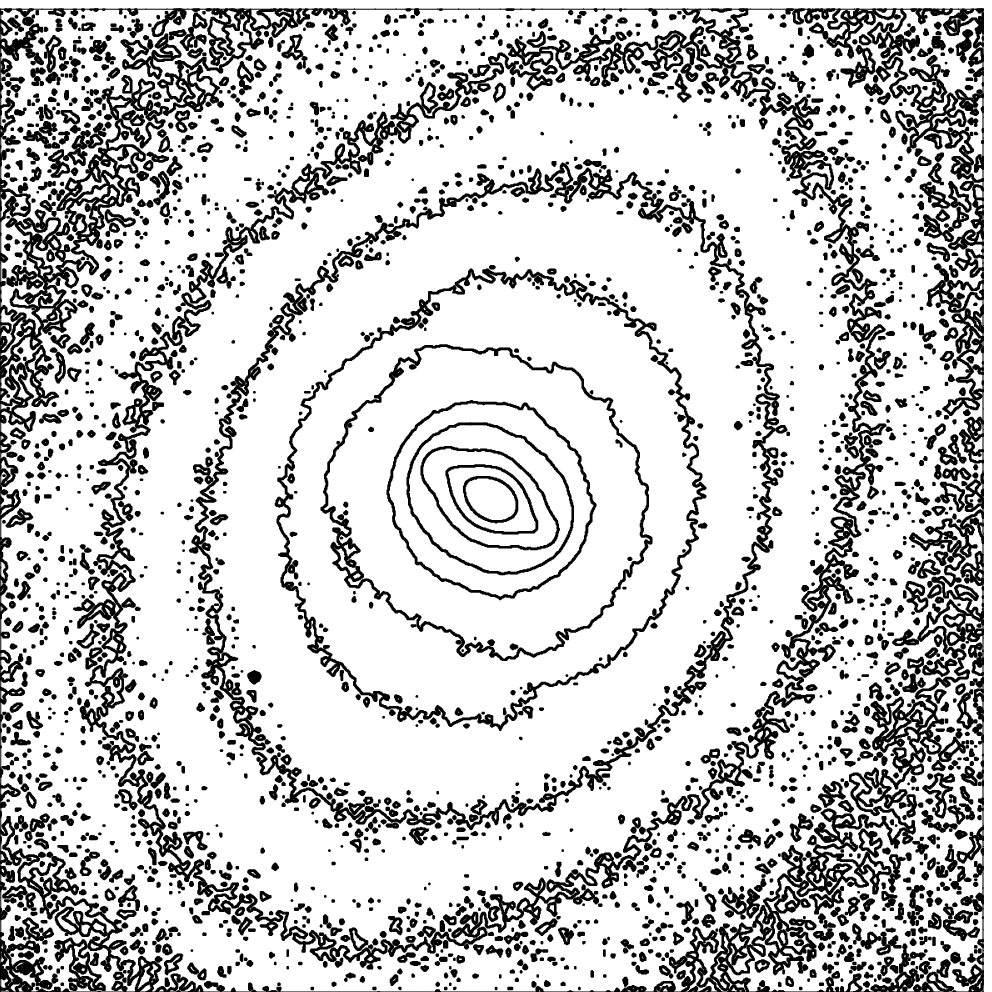}{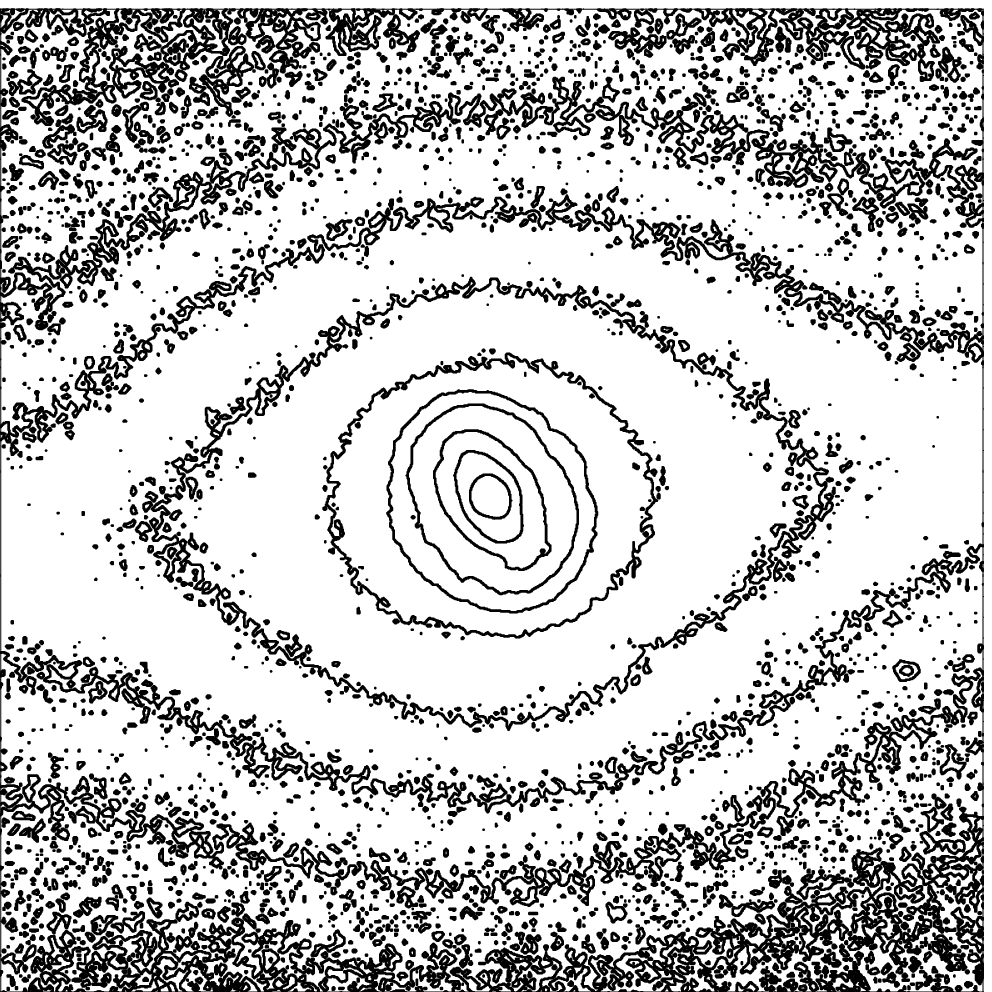}
\caption{I-band images of two galaxy prototypes with bars within bars,
  NGC\,1317 on the left, and NGC\,1433 on the right (from Wozniak et
  al. 1995).  The pixel size is 0.38\arcsec, and the seeing
  $\sim$1.1\arcsec.  The frames are 90\arcsec\ wide.  The contour
  scale is logarithmic with a spacing of 0.5\,mag.}
\end{figure}

\subsubsection{Dynamical Decoupling (Bars within Bars).}
Many disk galaxies host not only one primary large-scale bar but also
another misaligned secondary (nuclear) one, which is embedded (nested)
inside the primary bar (de Vaucouleurs 1974; Buta \& Crocker 1993;
Wozniak et al. 1995; Friedli et al. 1996; Jungwiert et al. 1997;
Mulchaey et al. 1997).  So far, the averaged length ratio between both
bars $a_p/a_s\approx7$, and primary bars are on average stronger than
secondary ones. But these values might be biased since actual
resolution only allows the largest secondary bars to be detected, and
tends to underestimate their ellipticities.  It is not yet clear if
secondary structures are thin bars or rather triaxial bulges.  The
angle between the two bars $\theta\!=\!\rm PA_p-PA_s$ does not take
any particular value.  This is one of the most decisive evidences of
the existence of a dynamical decoupling in the center of these
galaxies, i.e. primary and secondary bars should rotate at different
speeds.  Well-known objects include NGC\,1317 (Fig.~1,
$\theta\footnote{With respect to the direction of rotation, the
  secondary bar can either lead ($\theta > 0$), or trail the primary
  bar ($\theta < 0$).} \approx -88\deg$), NGC\,1433 (Fig.~1, $\theta
\approx +64\deg$), NGC\,5850 ($\theta \approx -65\deg$), and NGC\,4321
($\theta \approx 0\deg$, Knapen et al. 1995).  However, in the latter
case, it is not easy to determine if both bars are rotating with one
single pattern, or with two distinct patterns observed in this
particular configuration.

Orbits in galaxies with bars within bars have nicely been investigated
by Pfenniger \& Norman (1990) and Maciejewski \& Sparke (1997) using
non-evolving analytical potentials. However, self-consistent numerical
simulations remain the only way to study properly the formation and
evolution of such non-linear, non-axisymmetric, dissipative and
time-dependent systems.  Friedli \& Martinet (1993) showed that
embedded bars with $\Omega_s /\Omega_p \!>\! 1$ (i.e. $\theta \!=\!
\theta(t)$) can exist and be stable over many rotations.  The proper
way to proceed is to form first the primary bar which then accumulates
gas into the central region.  This allows ILRs to be present and to
shift gas forward.  Then, these great quantities of leading gas
naturally trigger the dynamical decoupling and the birth of the
secondary bar. A large fraction of the angular momentum continuously
lost by the gas is gained by the secondary bar, preventing it to
realign quickly.  Finally, the fact that CR$_s$ approximately
coincides with ILR$_p$ minimizes the possible negative effects of
resonances like the generation of too much chaos.

While large-scale bars appear to favor the central star formation
activity on certain condition (Hawarden et al. 1986; Martinet \&
Friedli 1997; Ho et al. 1997), they are not directly implicated in the
fueling of Active Galactic Nuclei (AGN) according to Ho et al. (1997).
In an inspiring paper, Shlosman et al. (1989) suggested that bars
within bars might rather play a key role in fueling AGN. In their
scenario, the secondary bar was a gaseous bar, not a stellar one.
While this possibility cannot be ruled out at all, the above-mentioned
observations and simulations correspond rather to another mechanism, a
``modified'' scenario where the gas fueling is driven by the secondary
stellar bar.  In a single bar, the scale of the compressed (fueled)
gas typically is $a_g \approx 0.1 a_p$. In the double-bar case, $a_g
\approx 0.1 a_s \approx 0.1 a_p/7 \approx 71$\,pc with $a_p\!=\!
5$\,kpc.  Note that large amounts of gas can be stacked along the
nuclear ring so that there is less, but more centrally concentrated,
gas in this case.  Although packing down gas that deep into the the
potential well is already impressive, there is still a long way to the
very center and the supermassive black hole. At some point, other
stages should follow, e.g. the formation of a tertiary bar, or the
Shlosman's mechanism.

So far, a total of $\sim$40 unambiguous double-barred
galaxies\footnote{I suggest to use simply the symbol S2B to denote
  these objects.}  have been discovered.  Nearly all have Hubble types
earlier than S2Bb, and $\sim$1/3 are known to be Seyferts. Of course,
this global sample is biased and does not allow to infer proper
statistics.  There are two nearly unbiased NIR surveys, but with only
arcsec resolution. The first one by Jungwiert et al. (1997) includes
72 galaxies. The authors found $\sim$24\% of S2Bs and $\sim$1/4 of
Seyferts. The other survey by Mulchaey et al. (1997) contains 30
Seyferts and 25 ``normal'' galaxies. They identified $\sim$16\% of
S2Bs but did not find any excess of double-bars in the sample of
Seyferts with respect to the non-Seyferts one (Mulchaey \& Regan
1997).

Clearly, bars within bars represent a very common phenomenon but seem
neither a necessary, nor a sufficient condition to create AGN. This
does not mean that such systems have nothing to do with central
activity. Indeed, the fueling is probably more episodic than
continuous. If all galaxies host a SBH in their center, then among
S2Bs the ratio of non-Seyferts to Seyferts is a direct measure of the
ratio of quiescent to active phases. The existence of Seyferts not of
the S2B type means either that the secondary bar is not yet resolved,
or other fueling mechanisms ($m\!=\!1$ mode; Shlosman et al. scenario)
are at play.

\begin{figure}[t]
\plotfiddle{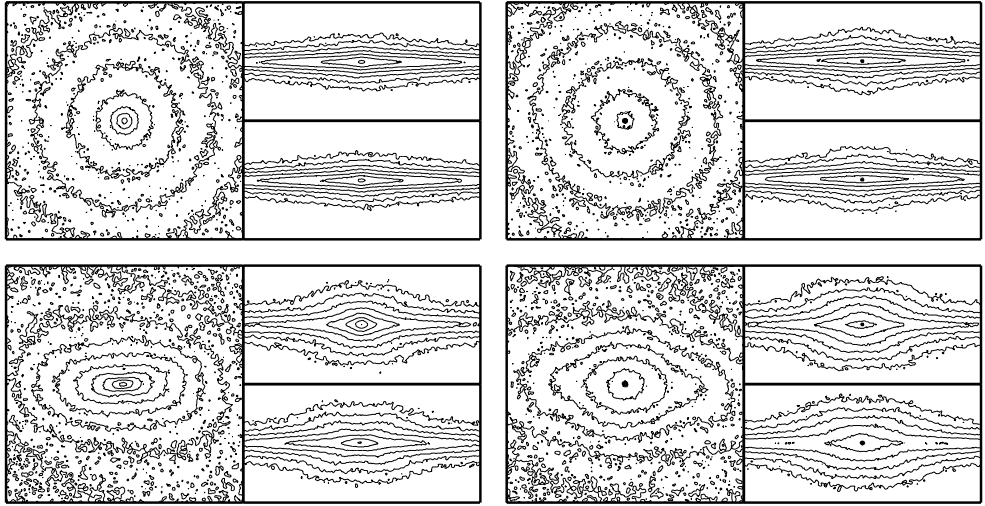}{65truemm}{0}{70}{70}{-210}{-35}
\caption{Projected views of some 3D self-consistent collisionless 
  numerical simulations with 720\,000 particles, and after an
  evolution of 2\,Gyr. {\it Top left.}  Purely axisymmetric model.
  {\it Bottom left.} Strongly barred model. {\it Top right.}
  Axisymmetric model with a central black hole with $M_{\rm
    bh}\!=\!0.02 M_*$. {\it Bottom right.} Weakly barred model
  including a 2\% black hole.}
\end{figure}

\subsubsection{Bar Thickening (Box-Peanut Formation).}
Only young bars are thin ($c/a\approx0.1$). After a few bar rotations,
typically 1\,Gyr, their inner parts (up to a few kpc) inflate and
become much thicker ($c/a\approx0.3$), being prone to the formation of
box- or peanut-shaped bulges (Fig.~2; Combes \& Sanders 1981; Combes
et al.  1990; Pfenniger \& Friedli 1991; Raha et al. 1991).  Berentzen
et al.  (1998) found that the box-peanut is much less pronounced if
the gas mass fraction is significant.  Many external galaxies harbor
such a central morphology when seen close to edge-on (Jarvis 1986;
Shaw et al. 1990; Bureau \& Freeman 1997), a property also shared by
the Milky Way (e.g. Dwek et al. 1995). This naturally leads to
establish a connection between bars and box-peanut bulges; this is
amply confirmed by the signatures of gas kinematics either on direct
(Kuijken \& Merrifield 1995), or on retrograde orbits (Emsellem \&
Arsenault 1997).

In the literature, this instability received a wide variety of names:
bending, box-peanut, buckling or fire-hose instability. There is still
some controversy about the nature of this instability: is it a
resonant bending fed by vertical diffusion of orbits (Combes et al.
1990; Pfenniger \& Friedli 1991) or a collective instability similar
to the fire-hose one (Raha et al. 1991; Merritt \& Sellwood 1994)?
The former possibility is due to the space part (vertical resonances)
of the distribution function (DF) while the latter is the result of
the velocity part of the DF (critical ratio of vertical to radial
velocity dispersions).  Unless unrealistically thin stellar disks are
built, this critical ratio cannot be reached at distances where
box-peanuts are formed, i.e. in rotation-dominated regions. But in
central, dispersion-dominated, areas, the fire-hose instability could
occur for instance within cold young stellar disks (Griv \& Chiueh
1998).

Thus, observed box-peanuts probably result from a 2/1/1 resonant
bending (ratio of vertical/radial/circular frequencies).  Before the
bending, the orbital structure is dominated by the 2D $x_1$ (direct)
and $x_4$ (retrograde) periodic orbit families.  After the bending,
fully 3D families appear, mainly the 2/1/1 banana and anti-banana ones
(bifurcation from the $x_1$) and the 1/1/1 anomalous ones (bifurcation
from the $x_4$).  At least two misconceptions about box-peanuts are
widespread: 1) {\it Box-peanuts end at $R_{\rm CR}$.} This is wrong!
Box-peanuts end near the vertical ILR, i.e. typically at $\sim$$R_{\rm
  CR}/2$.  2) {\it Box-peanuts require $z$-symmetry breaking to form.}
This is also wrong! While in many models, forcing the $z$-symmetry
indeed slows down the box-peanut growth rate by a factor of 3--4, it
is possible to find models where box-peanuts quickly appear without
any macroscopic $z$-asymmetry.

While the addition of a SBH amplifies the bulge inflating (something
only observed in barred models), this also highly alleviates or
suppresses the box-peanut shape (Fig.~2; Friedli 1994).  As a matter
of fact, in the bar region broad radial and vertical resonances are
unavoidable after the (dissipative) growth of a significant central
mass concentration. Stars are then allowed to diffuse into the bulge
from the disk (Pfenniger \& Norman 1990). The bulge-to-disk ratio
increases and secular evolution along the Hubble sequence constitutes
a natural outcome of this resonant heating.  Of course, the
inescapable formation of bar-driven bulges does not preclude other
processes to contribute as well. Some mass fraction of the bulges
could be primordial, and result from other secular evolution
mechanisms like minor mergers or Zhang's theory (Zhang 1999).

\begin{table}[t]
\caption{Summary of the bar-induced evolution of stellar and gaseous 
radial abundance profiles in different regions.} 
\vskip -5truemm
\begin{center}
\begin{tabular}{llc}
\tableline
\noalign{\vspace{0.8truemm}}
Regions & Stellar gradient & Gaseous gradient\\
        &                  & {\sc young bar} ($\la$1\,Gyr) 
              $\rightarrow$ {\sc old bar} ($\ga$1\,Gyr)\\
\noalign{\vspace{0.8truemm}}
\tableline
\noalign{\vspace{0.8truemm}}
 Bar        & Weak changes & Steep \ $\rightarrow$\  Flat\\
 Corotation & Plateau      & Break \ $\rightarrow$\  Flat\\
 Disk       & Flattening   & Flattening \ $\rightarrow$\  Flat\\
\noalign{\vspace{0.8truemm}}
\tableline
\tableline
\end{tabular}
\end{center}
\end{table}

\subsubsection{Abundance Profile Alteration.}
Outstanding properties of barred systems include the existence of
irregular (chaotic) orbits (e.g. Contopoulos \& Grosbøl 1989), and
gravitational torques. These characteristics lead both to large-scale
($\ga$1\,kpc) diffusion and mixing of stars, and produce transfer of
angular momentum and matter, especially gas (e.g. Athanassoula 1992).
This is not harmless for the stellar and gaseous abundance profiles,
as discussed in short below (for details see Friedli 1998).

$N$-body or orbit studies have shown that chaos increases if bar
strength, central mass concentration, noise, or asymmetries increases.
As shown by Pfenniger \& Friedli (1991), strong 3D $N$-body bars
typically host hot ($\sim$35\%, chaotic orbits), bar ($\sim$45\%), and
disk ($\sim$20\%) populations.  There is a significant diffusion in
both $R$ and $z$ directions as soon as the Lagrangian points $L_{4,5}$
become complex unstable which occurs at some critical bar strength
(Oll\'e \& Pfenniger 1998).  As a consequence, bar-generated features
appear in stellar abundance profiles (Friedli 1998).  When a strong
bar appears, any smooth and steep initial abundance gradient $d\log
A/dR\!<\!0$ is then distorted by a plateau near $R_{\rm CR}$ and a
pronounced flattening in the disk region.  Models with $d\log
A/dz\!<\!0$ quickly develop box-peanut shaped isoabundance contours,
whereas models with $d\log A/dz\!=\!0$ become X-shaped and present
positive gradient in the disk region.

Bar-generated features also appear in radial gaseous abundance
profiles: in the disk region a severe flattening occurs (Friedli et
al. 1994) and a break is observed near $R_{\rm CR}$ in young bars
(Martinet \& Friedli 1997).  Indeed, in the bar region during the
early phase of its existence, the gradient is maintained since the gas
dilution by significant gas inflow is compensating for the
heavy-element production by star formation.  The oxygen abundances
from H\,{\sc ii} regions in several late-type galaxies have clearly
revealed that barred galaxies have weaker gradients (Vila-Costas \&
Edmunds 1992). Moreover, Martin \& Roy (1994) showed that the stronger
the bar, the shallower the gradient. Breaks in abundance profiles have
been observed in NGC\,3359 (Martin \& Roy 1995), and NGC\,1365 (Roy \&
Walsh 1997) suggesting that these bars are of the order of 1\,Gyr old.

Table~1 summarizes the bar-induced evolution of stellar and gaseous
radial abundance profiles in different regions.

\subsection{Death}
Most bars are probably not perpetual. They might be dissolved by
internal or external processes, but the basic reason is the same:
significant central mass concentration is always fatal.  It generates
ILRs which strongly modifies the orbital structure. The bar-supporting
$x_1$ orbits are depopulated in favor of chaotic and anti-bar $x_2$
orbits; the bar cannot survive (Pfenniger \& Norman 1990; Hasan \&
Norman 1990).  Possible mechanisms of central mass accumulation
include bar-driven gas accretion, one could speak of a suicide
(Friedli \& Benz 1993; Norman et al. 1996; Berentzen et al.  1998),
growth of SBHs (Friedli 1994), accretion of satellites (Pfenniger
1991), or the extreme case of mergings (Barnes \& Hernquist 1991).
Massive nuclear rings also act to weaken or dissolve the bar, mainly
exterior to the ring (Heller \& Shlosman 1996).

The critical threshold to annihilate the bar can be defined as
$\gamma_{\rm anni} \equiv M_{\rm center}/M_{\rm total}$.  The bar
annihilation time-scale corresponds to the time necessary to reach
$\gamma_{\rm anni}$.  It depends on the accretion rate, and probably
spans a wide interval, something like 0.2--20\,Gyr.  In fact,
$\gamma_{\rm anni}$ depends on both mass and concentration.  This has
not always been underlined properly leading to widely different
numbers in the literature, $\gamma_{\rm anni} \approx 1-10\%$.  A more
accurate but still rough assessment is:
\begin{equation} 
\gamma_{\rm anni} \approx 0.1 R + 0.02 
\quad \quad \quad 0 < R \, [\rm kpc] \le 1 \, ,
\end{equation} 
where $R$ is the radius considered. SBHs are especially efficient bar
annihilators with $\gamma_{\rm anni} \approx 0.02$ (Friedli 1994).  A
similar value is found for the deletion of triaxiality in ellipticals
(Merritt \& Quinlan 1998).  Note that the strongest bars are the most
sensitive to the addition of compact mass in their cores; they quickly
become round in the central region, long before $\gamma_{\rm anni}$ is
effectively reached.

Demographic studies of SBHs find a correlation between the mass of the
black hole and the one of the bulge (or hot) component (Kormendy \&
Richstone 1995; Magorrian et al.  1998), $M_{\rm bh} \approx 0.005
M_{\rm bulge}$. The scatter is large, with late-type objects having
lighter SBHs than early-type ones.  The upper limit seems roughly
$M_{\rm bh} \la 0.025 M_{\rm bulge}$, surprisingly similar to
$\gamma_{\rm anni}$. This might indicate that bars play a role in the
SBH growth which is then turned off as soon as the bar has
disappeared.

In Sect.~2.2, we have seen that a bar plus a SBH are able to generate
a significant bulge. Dissolved bars are thus to be searched for in
unbarred early-type galaxies, i.e. S0/Sa's.  The most massive SBHs
should be concealed in those objects as well.  The proof of the
existence of dissolved bars in early-type galaxies seem to have been
established by Dutil \& Roy (1999) through an abundance study.
Indeed, they found that the extrapolated central O/H abundances of
early-types (both barred and unbarred) are very similar, whereas that
of barred late-types is systematically $\sim$0.5\,dex lower than the
unbarred ones.

\section{Gaseous Bars}
A bar-like morphology for the gas is present within some stellar bars
(e.g. NGC\,7479, Laine et al. 1999).  Numerical simulations have shown
that young and strong stellar bars without ILRs can host large-scale
gaseous and ``H$\alpha$'' bars (Martin \& Friedli 1997).  Such {\it
  induced gaseous bars} generally lead the stellar bar by a few
degrees, $a_g \!<\! a_*$, and $(b/a)_g \!<\! (b/a)_*$.  Typically, the
total gas mass $M_g \la 10^9 \, \rm M_{\odot}$, and the (nearly
constant) maximum gas surface density $\Sigma_g^{\rm max} \approx 2.5
\cdot 10^3 \, \rm M_{\odot} \, pc^{-2}$. With time, both $a_g$ and
$M_g$ first decrease very quickly, and then reach an asymptotic value.
The evolution of $M_g$ and $\Sigma_g^{\rm max}$ are essentially
controlled by the self-regulated star formation processes, not by
dynamics.  This might explain why generally $M_g \la 0.3 M_{\rm dyn}$,
where $M_{\rm dyn}$ is the dynamical mass.  The gaseous bar lifetime
is of the order of 1\,Gyr.  For bars within bars systems, a gaseous
bar might be present inside the secondary bar as well, but with
length, mass, and lifetime scale down by a significant factor.

The galaxy NGC\,6946 somewhat represents an issue for this process of
stellar bar-driven gaseous bar: it has a gaseous bar, but no (or very
weak) stellar bar (Regan \& Vogel 1995).  Hence, one could imagine
that {\it spontaneous gaseous bar} instability might occur in some
galaxies where significant gas self-gravitation ($M_g \ga 0.3 \,
M_{\rm dyn}$) is present.  Such a critical mass could be reached in
primordial galactic discs, mergers, and of course stellar bars.
However, spontaneous gaseous bars are very unstable (fragmentation),
and gas self-gravity decreases quickly due to furious star formation.
A very short lifetime results.  Another appealing possibility for
NGC\,6946 is that the gaseous bar had actually been formed by a strong
stellar bar which is now nearly dissolved.

Gaseous bars can also be found in relatively unexpected places, e.g.
at the center the giant elliptical Centaurus~A (Mirabel et al. 1999).
This galaxy harbors an AGN and relativistic jets powering two
spectacular large lobes separated by $\sim$350\,kpc.  If this
$\sim$5\,kpc long bi-symmetric structure (gaseous bar plus spiral
arms) serves to fuel the AGN, its lifetime should have been long
enough to produce the radio lobes ($\ga$10\,Myr).  In the Sb galaxy
Circinus, Maiolino et al. (1999) have also found a $\sim$100\,pc
nuclear gas bar possibly feeding an AGN.  Both gaseous bars seem not
to have distinct stellar counterparts.

\begin{table}[b]
\caption{Summary of the main differences between young and old bars.} 
\vskip -5truemm
\begin{center}
\begin{tabular}{lll}
\tableline
\noalign{\vspace{0.8truemm}}
 Properties & Young stellar bars    & Old stellar bars \\
            & (less than 1--2\,Gyr) & (more than 1--2\,Gyr) \\
\noalign{\vspace{0.8truemm}}
\tableline
\noalign{\vspace{0.8truemm}}
 Thickness           & Thin      & Thick (box-peanuts) \\
 Pattern speed       & High      & Low \\
 Central decoupling  & Unlikely  & Likely (S2Bs) \\
 Radial abundance profile (gas) & Break near $R_{\rm CR}$ & Flat \\
 Star formation $^a$ & Intense   & Moderate \\
                     & Along bar $\rightarrow$ Center & Circumnuclear ring \\
\noalign{\vspace{0.8truemm}}
\tableline
\tableline
\end{tabular}
\end{center}
\vskip -2truemm
$^a$ Not discussed here; see Martin \& Friedli (1997)
\vskip +2truemm
\end{table}

\section{Summary and Conclusion}
(Galaxies with) bars evolve morphologically, dynamically, chemically,
at small- and large-scales, over short and long timescales.  Some of
their properties change and it is then possible to distinguish young
bars from old ones as summarized in Table~2.  The bar-induced galaxy
evolution along the Hubble sequence definitely tends to transform
late-type objects into early-type ones.  Clearly, like living beings,
bars are born, age, and die!  But many fascinating riddles still
subsist:\\
-- Could some stellar bars be ``eternal'', i.e. be robust over many
Hubble times?\\
-- Could stellar bars ``revive'', i.e. be recurrent? \\
-- At what $z$ does the first bar appear?\\
-- At what $z$ does $\beta$ reach its maximum?\\
Although the life cycle of bars is not yet fully understood, it should
certainly bear some resemblance with the one depicted in Fig.~3\dots

\begin{figure}[t]
\plotfiddle{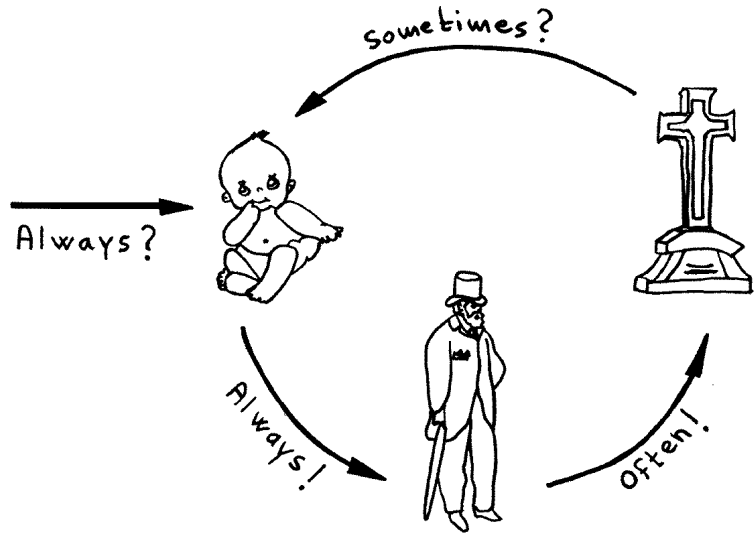}{60truemm}{0}{140}{140}{-140}{-15}
\caption{Humoristic illustration of the possible life cycle of bars.}
\end{figure}

\acknowledgments 
For so many enlightening and lively discussions on bars over the past
twelve years, I am very indebted to L.\,Athanassoula, J.\,Beck\-man,
W.\,Benz, F.\,Combes, S.\,Courteau, Y.\,Dutil, E.\,Emsellem, R.\,Fux,
H.\,Hasan, C.\,Heller, L.\,Ho, S.\,Jogee, J.\,Kenney, R.\,Kennicutt,
J.\,Knapen, S.\,Laine, P. Martin, L.\,Martinet, D.\,Merritt,
M.\,Noguchi, C.\,Norman, D.\,Pfenniger, J.-R.\,Roy, J.\,Sellwood,
M.\,Shaw, I.\,Shlosman, S.\,Udry, K.\,Wada, H.\,Wozniak, and many more
I am certainly forgetting!


\end{document}